\documentclass[aps,twocolumn,showpacs,nofootinbib]{revtex4}
\usepackage{graphicx}

\newcommand{\beq}{\begin{equation}}
\newcommand{\eeq}{\end{equation}}
\newcommand{\beqa}{\begin{eqnarray}}
\newcommand{\eeqa}{\end{eqnarray}}

\def\half{\frac{1}{2}}

\def\half{\frac{1}{2}}

\def\<{\langle}
\def\>{\rangle}

\newcommand{\complex}{{\kern .1em {\raise .47ex\hbox {$\scriptscriptstyle |$}}\kern -.4em {\rm C}}}
\newcommand{\real}{{{\rm I} \kern -.19em {\rm R}}}

\begin{document}

\title{Can relativity be considered complete ?\\
From Newtonian nonlocality to quantum nonlocality and beyond}

\author{Nicolas Gisin}

\affiliation{
    Group of Applied Physics, University of Geneva, 1211 Geneva 4,
    Switzerland}

\date{\today}

\begin{abstract}
We review the long history of nonlocality in physics with special emphasis on the conceptual
breakthroughs over the last few years. For the first time it is possible to study "nonlocality
without signaling" {\it from the outside}, that is without all the quantum physics Hilbert space
artillery. We emphasize that physics has always given a nonlocal description of Nature, except
during a short 10 years gap. We note that the very concept of "nonlocality without signaling" is
totally foreign to the spirit of relativity, the only strictly local theory.
\end{abstract}

\maketitle

\section{Introduction}\label{intro}
100 years after Einstein miraculous year and 70 years after the EPR paper \cite{EPR35}, I like to
think that Einstein would have appreciated the somewhat provocative title of this contribution.
However, Einstein would probably not have liked its conclusion. But who can doubt that relativity
is incomplete? and likewise that quantum mechanics is incomplete! Indeed, these are two scientific
theories and Science is nowhere near its end (as a matter of fact, I do believe that there is
no end \cite{Weinberg}). Well, actually, I am, of course, not writing for Einstein, but for those
readers interested in a (necessarily somewhat subjective) account of the {\it peaceful
co-existence} \cite{peacefulCoexistence} between relativity and quantum physics in the light of the
conceptual and experimental progresses that happened during the last ten years, set in the broad
perspective of physics and nonlocality since Newton \cite{Gilder05}.

\section{Non-locality according to Newton}\label{Newton}
Isaac Newton, the great Newton of Universal Gravitation, was not entirely happy with his theory.
Indeed, he was well aware of an awkward consequence of his theory: if a stone is moved on the moon,
then our weight, of all of us, here on earth, is {\it immediately} modified. What troubled so much
Newton was this {\it immediate} effect, i.e. the nonlocal prediction of his theory. Let's read how
Newton described it himself \cite{NewtonNonlocality}:

{\it That Gravity should be innate, inherent and essential to Matter, so that one Body may act upon
another at a Distance thro' a Vacuum, without the mediation of any thing else, by and through which
their Action and Force may be conveyed from one to another, is to me so great an Absurdity, that I
believe no Man who has in philosophical Matters a competent Faculty of thinking, can ever fall into
it. Gravity must be caused by an Agent acting constantly according to certain Laws, but whether
this Agent be material or immaterial, I have left to the Consideration of my Readers. }

It would have been hard for Newton to be more explicit in his rejection of nonlocality! However,
most physicists didn't pay much attention to this aspect of Newtonian physics. By lack of
alternative, physics remained nonlocal until about 1915 when Einstein introduced the world to
General Relativity. But let's start ten years earlier, in 1905.

\section{Einstein, the greatest mechanical engineer}\label{EinsteinEngineer}
In 1905 Einstein introduced three radically new theories or models in physics. Special relativity
of course, but more relevant to this section are his descriptions of Brownian motion and of the
photo-electric effect. Indeed, both descriptions show Einstein's deep intuition about mechanics.
Brownian motion is explained as a complex series of billiard-ball-like-collisions between a visible
molecule - the particle undergoing Brownian motion - and invisible smaller particles. The random
collisions of the latter explaining the erratic motion of the former. Likewise, the photo-electric
effect is given a mechanistic explanation. Light beams contain little billiard-balls whose energy
depends on the color, i.e. wavelength, of the light. These light-billiard-balls (today called
photons and recognized as not at all billiard-ball-like) hit the electrons on metallic surfaces and
mechanically kick them out of the metal, provided they have enough energy.

General relativity can also be seen as a mechanical description of gravitation. When a stone is
moved on the moon, a bunch of gravitons (in modern terminology) fly off in all directions at a
finite speed, the speed of light. Hence, about a second later, the earth is {\it informed} and only
then is our weight affected. This is, I believe, the greatest achievement of Einstein, the greatest
mechanical engineer\footnotemark\footnotetext{My friends know well that in my mouth "engineer" has
no negative connotation, quite the opposite. For me, a physicist must be a good theorist and a good
engineer! Well, I warned you, dear reader, this is a somewhat subjective article.} of all times:
{\bf Einstein turned physics into a local theory!}

\section{Quantum mechanics is not mechanical}\label{Qphysics}
Only about ten years after general relativity came quantum mechanics. This was quite an
extraordinary revolution. Until then, greatly thanks to Newton and Einstein's genius, Nature was
seen as made out of many little {\it billiard-balls} that mechanically bang into each other. Yet,
quantum mechanics is characterized by the very fact that it no longer gives a mechanical
description of Nature. The terminology quantum {\it mechanics} is just a historical mistake, it
should be called {\it Quantum Physics} as it is a radically new sort of physical description of
Nature.

But this new description let nonlocality back into Physics! And this was unacceptable for Einstein.

It is remarkable and little noticed that since Newton, physics gave a local description of Nature
only during some 10 years, between about 1915 and 1925. All the rest of the time, it was nonlocal,
though, with quantum physics, in quite a different sense as with Newton gravitation. Indeed, the
latter implies the possibility of arbitrarily fast signaling, while the former prohibits it.

\section{Non-locality according to Einstein}\label{Einstein}
In 1935 two celebrated papers appeared in respectable journals, both with famous authors, both
stressing the - unacceptable in their authors view - nonlocal prediction of quantum physics
\cite{EPR35,Schrodinger35}. A lot has been written on the EPR "paradox" and I won't add to this. I
believe that Einstein's reaction is easy to understand. Here is the man who turned physics local,
centuries after Newton wrote his alarming text, he is proud of his achievement and certainly
deserves to be. Now, only a few years latter, nonlocality reappears! Today one should add that
quantum nonlocality is quite a different concept from Newtonian nonlocality, but Einstein did
not fully realize this.

What Einstein and his colleagues saw is that quantum physics describes spatially separated
particles as one global system in which the two particles are not logically separated. What they
did not fully realize is that this does not allow for signaling, hence it is not in direct conflict
with relativity. In the next section I'll try to present this using modern terminology.

Most physicists didn't pay much attention to this aspect of quantum physics. A kind of
consensus established that this was to be left for future examination, once the technology would be
more advanced. The general feeling was that quantum nonlocality was nothing but a laboratory
curiosity, not serious physics.

Young physicists may have a hard time to believe that such an important concept, like quantum
nonlocality, was, during many decades, not considered as serious. But this was indeed the real
state of affairs: ask any older professors, a vast majority of them still believes that it is
unimportant. Let me add two little stories that illustrate what the situation was like. John Bell,
the famous John Bell of the Bell inequalities and of the Bell states, never had any quantum physics
student. When a young physicist would approach him and talk about nonlocality, John's first
question was: "Do you have a permanent position?". Indeed, without such a permanent position it was
unwise to dare talking about nonlocality! Notice that John Bell almost never published any of his
remarkable and nowadays famous papers \cite{BellSpeakable} in serious journals: the battle with
referees were too ... time wasting (not to use a more direct terminology). Further, if you went to
CERN where John Bell held a permanent position in the theory department and asked at random about
John's contributions to physics, his work on the foundation of quantum physics would barely be
mentioned (true enough, he had so many other great
contributions!)\footnote{Another story happened to me while I was a young post-doc
eager to publish some work. In a paper \cite{GisinJMP83} I wrote "A quantum particle may disappear
from a location A and simultaneously reappear in B, without any flow in-between". The referee
accepted the paper under the condition that this outrageous sentence is removed. This referee
considered his paternalist attitude so constructive that he declared himself to me: "look how
helpful I am to you" (admittedly, he was politically correct).}.

Anyway, so quantum nonlocality remained for decades in the {\it curiosity lab} and no one
paid much attention. But in the 1990's two things changed. First, a conceptual breakthrough
happened thanks to Artur Ekert and to his adviser David Deutsch \cite{Ekert91}. They showed that
quantum nonlocality could be exploited to establish a cryptographic key between two distant
partners and that the confidentiality of the key could be tested by means of Bell's inequality.
What a revolution! This is the first time that someone suggested that quantum nonlocality is not
only real, but that it could even be of some use. A second contribution came from the progress in
technology. Optical fibers had been developed and installed all over the world. And Mandel's group
at the University of Rochester (where I held a one-year post-doc position and first met with
optics) applied parametric down-conversion to produce entangled photon pairs \cite{Mandel}. This
was enough (up to the detectors) to demonstrate quantum nonlocality outside the curiosity
laboratory. In 1997 my group at Geneva University demonstrated the violation of Bell inequalities
between two villages around Geneva, see Fig. 1, separated by a little more than 10 km and linked by a
15km long standard telecom fiber \cite{Tittel98,Tittel99} (since then, we have achieved 50km
\cite{Marcikic04}). So quantum nonlocality became politically acceptable! But what is it? Let me
introduce the concept using students undergoing "quantum exams".

\section{Quantum exams: entanglement}\label{Qexams}
Assume that two students, Alice and Bob, have to pass some exams. As always for exams, the
situation is arranged in such a way that the students can't communicate during the exam. Clearly however,
they are allowed, and even encouraged, to communicate beforehand. Alice and Bob know in
advance the list of possible questions, they also know that this is a kind of exam allowing only
a very limited number of possible answers, often only a binary choice between {\it yes} and {\it
no}. During the exam Alice receives one question out of the list, let's denote it by $x$; Bob receives
question $y$. Finally, denote $a$ and $b$ Alice and Bob's answers, respectively. Hence, an exam is
a realization of a random process described by a conditional probability function, often merely
called a {\it correlation}:
\beq
P(a,b|x,y)
\eeq
Clearly, the choice of questions $x$ and $y$ are under the professor's control. However, as all
professors know, the students' answers $a$ and $b$ are not! This is similar to experiments: the
choice as to which experiment to perform is under the physicists control, but not the answer given by
Nature.

In the following, we shall consider three kinds of exams, in order to understand what kind of
constraints they set on the correlation $P(a,b|x,y)$.

\subsection{Quantum exam \#1}\label{QE1}
In this first kind of quantum exam Alice is asked to tell which question is given to Bob, and
vice-versa. This is clearly an unfair exam! Why? Because Alice and Bob are not supposed to
communicate. How could they then succeed with a probability greater than mere
chance\footnotemark\footnotetext{Somewhat surprisingly there is a strategy such that the
probability that both players succeed is 50\%.}? This simple example shows that prohibiting
signaling already limits the set of possible correlation $P(a,b|x,y)$. For example
$P(a,b|x,y)=\delta(a=y)\delta(b=x)$ is excluded. Notice that a correlation $P(a,b|x,y)$ is
non-signaling if and only if its marginal probabilities are independent of the other side input:
$\sum_b P(a,b|x,y)$ is independent of $y$ and $\sum_a P(a,b|x,y)$ is independent of $x$.

\subsection{Quantum exam \#2}\label{QE2}
The second kind of quantum exam is closer to standard exams. Alice and Bob are simply requested to
provide the same answer whenever they receive the same question. This is clearly feasible: we
all expect that good students give the same answer to the same question. It suffices that they
prepare for the exam well enough. Note that the quantum exam \#2 under consideration here is even
easier than standard exams, as there is no notion of correct or incorrect answers. All that is
required is that Alice and Bob give consistent answers: it suffices that they jointly decide in
advance which answer to give for each of the possible questions.

Now, a central problem: Could Alice and Bob succeed with certainty for such an exam \#2 by other
means, that is without jointly deciding the answers in advance? Think about it. If you found an
alternative trick, then, if you are a student, you should use your trick to pass the next
examination: just apply your trick together with the best student, you'll get the same mark as
him/her\footnotemark\footnotetext{Admittedly, the danger is that both student that get the bad
mark! But, on average, the poor student improves.}. And if you are a professor and found a trick,
then you should stop testing your student with standard exams! Well, of course, there is no other
trick, at least none applicable to classical students.

Correlations that satisfy $P(a=b|x=y)=1$ are necessarily of the form
\beq
P(a,b|x,y)=\sum_\lambda
\rho(\lambda) Q(a|x,\lambda)Q(b|y,\lambda) \label{lhv}
\eeq
for some probability function $Q$ and some distribution $\rho$ of common strategy $\lambda$.
Historically, the $\lambda$ were called "local hidden variables", computer scientist call them
"shared randomness"; here the $\lambda$ denote common strategies.

$P(a=b|x=y)=1$ is but one example of a local correlation, among infinitely many others. Relation
(\ref{lhv}) characterizes all local correlations.

In summary, some exams require common strategies;
in other words, some observed correlations can't be explained except by common causes.

\subsection{Quantum exam \#3}\label{QE3}
The third kind of quantum exam is the most tricky and interesting. For (apparent) simplicity let's
restrict the set of questions and answers to binary sets and let us label them by bits, "0" and
"1". In this exam Alice and Bob are required to always output the same answer, except when they
both receive the question labelled "1" in which case they should output different answers. Note
that formally this exam requires that Alice and Bob's data satisfy the following equality, modulo
2: $a+b=x\cdot y$. This time it is not immediately obvious whether they can prepare a strategy that
guaranties success.

Assume first that the strategy forces Alice to output an answer that depends only on her input
$x$, i.e. Alice's strategy is deterministic. But in such a case, whenever Bob receives the question
1, he can't decide on his output since it should depend on Alice's question. Next, if Alice's
output is random, this is clearly of no help to Bob. Consequently there is no way for Alice and Bob
to succeed with certainty.

Let us emphasize that successfully completing this exam does not necessarily imply communication between Alice
and Bob. Indeed, assume that somehow Alice and Bob's data would always satisfy $a+b=x\cdot y$.
Would this allow Alice to communicate to Bob, or vice-versa? Well, it depends! If Alice's output
$a$ is known to Bob, for instance they decide on $a=0$ always, then whenever Bob receives $y=1$, he
can deduce Alice's question from $a+b=x\cdot y$ and from his output: $x=b$ in the example. But if
Alice's outcome is unknown to Bob, for instance if Alice outcome is merely a random bit, then the
relation $a+b=x\cdot y$ is of no help to Bob. We shall come back to this concept of a non-signaling
correlation satisfying $a+b=x\cdot y$ in section \ref{nonlocality}.

Let us define the mark $M$ of this quantum exam \#3 as the sum of the success probabilities
\cite{CHSH}:
\beqa
M&=&P(a+b=xy|x=0,y=0) \nonumber\\
&+&P(a+b=xy|x=0,y=1) \nonumber\\
&+&P(a+b=xy|x=1,y=0) \nonumber\\
&+&P(a+b=xy|x=1,y=1)
\eeqa
It is not difficult to realize that the optimal strategy for Alice and Bob consists in deciding in
advance on a common answer, independent of the questions they receive. With such a strategy they
are able to achieve the mark $M=3$. This is indeed the optimal mark achievable by common
strategies:
\beq
M\le3  \label{BellM}
\eeq
This is an example of a Bell inequality: a constraint on correlations arising from common
strategies. Interesting Bell inequalities are those that can be violated by quantum physics. In the
case of (\ref{BellM}), if Alice and Bob share singlets, then they can obtain the mark
$M_{QP}=2+\sqrt{2} \approx 3.41$. Tsirelson proved that this is the highest mark achievable using
quantum correlations \cite{TsirelsonBound}.

Accordingly, quantum theory predicts that some tasks can be achieved that can't be predicted by any
local mechanical model, i.e. some exams are passed with higher marks than classically possible. The
fact that such tasks were invented for the purpose of showing the superiority of quantum physics
doesn't affect the conclusion. Still, it is only once some useful and natural tasks were found,
concretizing the superior power of quantum physics over all possible local strategies, that quantum
nonlocality became accepted by the physics community\footnotemark\footnotetext{I wish someone
establishes the statistics of the occurrences of the words "Bell inequality" and "nonlocality" in
Physical Review Letters. I bet that a phase transition happen in the early 1990's, after Ekert's
paper on quantum cryptography. In 1997 I started a PRL with the sentence: "Quantum theory is
nonlocal." and got considerable reactions to what was felt as a provocative statement; today the
same statement can be found in many papers, not provoking any reaction.}.

\section{Coin tossing at a distance}\label{CoinTossing}
Another way to present nonlocality to non-physicist friends is the following. Imagine two
hypothetical players that toss coins. The players are separated in space and toss their coin once
per minute. They use their free-will to decide, for each toss, whether to use their right hand or
their left hand, independently of each other. And they mark all results (time, hand and head/tail)
in a big black laboratory notebook, see Fig. 1.

\begin{figure}
  \includegraphics
  [width=\columnwidth]{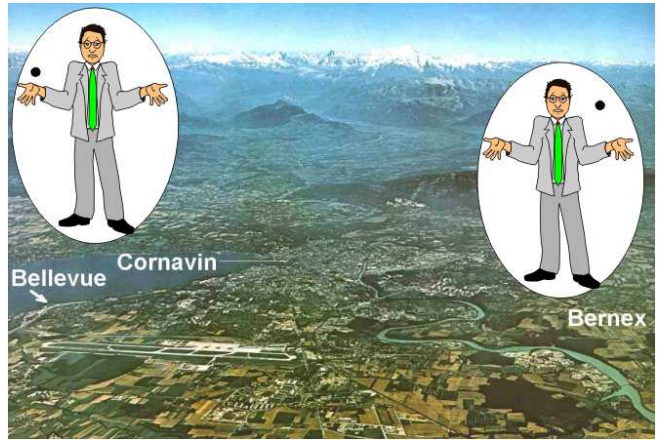}\\
  \caption{Bernex and Bellevue are the two villages north and south of Geneva between which our long-distance
  test of Bell inequality outside the lab was performed in 1997, section \ref{Einstein}. The inset represent
  two player that toss coins, as explained in section \ref{CoinTossing}. In the real experiment the coins were
  replaced by photons, the players by interferometers, their right and left hands by phase modulators and head/tail
  by two detectors. The experimental results are similar to that of the game, with weaker but still nonlocal correlations.}
\end{figure}

After thousands of tosses, they get bored. Especially, given that nothing interesting happens: for each of
the two players, heads and tails occur with a frequency of 50\%, independently of which hand they
use. Hence, the players decide to go for a beer. There, in the bar, they compare their notes and
get very excited. Indeed, quickly they notice that whenever at least one of them happened to have
chosen his right hand for tossing his coin, both players always obtained the same result: either
both head or both tails. But whenever, by mere chance, they both chose the left hand, then they
always obtained opposite results: head/tail or tail/head. A very remarkable correlation!

The observation of correlations and the development of theoretical models explaining them is the essence of the
scientific method. This is true not only in physics, but also in all other sciences, like geology
and medicine for instance. John Bell used to say "correlations cry out for explanations!" \cite{BellCorrCry}.

So, why are our two players that excited by the correlation they observe? Note that locally,
nothing interesting happens; in particular there is no way for one player to infer from his data
which hand the other player chose. Even if one player decides to always use the same hand, this
has no effect on the statistics observed by his colleague. Consequently, this game and the observed
correlation do not imply any signaling. So, why do we feel that this is impossible? Actually,
frankly, I do not know!

Classical correlations are always explained by either of two kinds of causes. The first kind is
"signaling", one player somehow informs or influences the other player. This is clearly not the
case here, since we assumed the players were widely separated in space (for the physicists we may add
"space-like separated"). The second kind of causes for classical correlations is a common cause.
For example all football players simultaneously stop running, because the umpire whistled. This
kind of cause is precisely equivalent to the assumption of a common strategy, as formalized by
(\ref{lhv}) and excluded for the present correlation by Bell's inequality (\ref{BellM}).
Consequently, the correlation observed by our two players is of a different nature. The big
surprise is that anything beyond the two classical causes for correlation exists! This is what
Einstein and many others had a hard time to believe. But, today, if one accepts this as a matter
of theoretical prediction and experimental confirmation, then the next big question is "why can't
the correlation observed by our hypothetical players not be observed in the real world?". Indeed,
quantum physics (and tensor products of Hilbert spaces) tell us that Bell's inequality
(\ref{BellM}) can be violated, i.e. not all quantum correlations can be explained by one of the two
kinds of classical causes for correlations, but quantum physics does not allow correlations as
strong as observed by our hypothetical players. Still, this game is illustrative of quantum nonlocality, as we shall elaborate in section \ref{nonLocality}

\section{Experiments: God does play dice, he even plays with nonlocal dice}\label{exp}
Physics is an experimental science and experiments have again and again supported the nonlocal
predictions of quantum theory. All kind of experiments have been performed, in laboratories
\cite{BellExpLab} and outside \cite{Tittel98,Tittel99,Weihs98}, with photons and with massive
particles \cite{Wineland01}, with independent observers to close the locality loophole
\cite{AspectLocLoophole,Weihs98,Tittel98,Tittel99,ZbindenSwitch}, with quasi-perfect detectors
\cite{Wineland01} to close the detection loophole, with high precision timing to bound the speed of
hypothetical hidden communication \cite{SpeedQI}, with moving observers to test alternative models
\cite{MovingObs} (multi-simultaneity \cite{SuarezScarani97} and Bohm's pilot wave
\cite{BohmPilotWave})\footnotemark\footnotetext{The conclusion that follows from all these
experiments is so important for the physicist's world-view, that an experiment closing
simultaneously both the locality and the detection loophole is greatly needed.}. All these results
proclaim loudly: {\bf God plays dice}. Note how ironic the situation is: the conclusion "God plays
dice" is imposed on us by the experimental evidence supporting quantum nonlocality and by
Einstein's postulate that no information can travel faster than light. Indeed, as mentioned in
sub-section \ref{QE3}, a violation of (\ref{BellM}) with deterministic outputs leads to signaling.
Consequently, the experimental violation of (\ref{BellM}) and the no-signaling principle imply
randomness \cite{PR94,PR9597}.

Actually, the situation is even more interesting: Not only does God play dice, but he plays with
nonlocal dice! {\bf The same randomness manifests itself at several locations}, approximating
$a+b\approx x\cdot y$ better than possible with any local classical physics model.

A very small minority of physicists still refuse to accept quantum nonlocality. They ask (sometimes
with anger) {\it How can these two space-time locations, out there, know about what happens in each
other without any sort of communication ?}. I believe that this is an excellent question! I have slept
with it for years \cite{HowCorrelation}. I summarize my conclusion in the next section.

\section{Entanglement as a cause of correlation}\label{ent}
Quantum physics predicts the existence of a totally new kind of correlation that will never have
any kind of mechanical explanation. And experiments confirm this: Nature is able to produce the
same randomness at several locations, possibly space-like separated. The standard explanation is
"entanglement", but this is just a word, with a precise technical definition
\cite{Werner88,Terhal03}. Still words are useful to name objects and concepts. However, it remains
to understand the concept. Entanglement is a new explanation for correlations. Quantum correlations
simply happen, as other things happen (fire burns, hitting a wall hurts, etc). Entanglement appears
at the same conceptual level as local causes and effects. It is a primitive concept, not reducible
to local causes and effects. Entanglement describes «correlations without correlata»
\cite{MerminIthaca} in a holistic view \cite{Esfeld04}. In other worlds, {\bf a quantum correlation
is not a correlation between 2 events, but a single event that manifests itself at 2 locations}.

Are you satisfied with my explanation of what entanglement is? Well, I am not entirely! But what is
clear is that entanglement exists. Moreover, entanglement is incredible robust! The last point
might come as a surprise, since it is still often claimed that entanglement is as elusive as a
dream: as soon as you try to talk about it, it evaporates! Historically this was part of the
suspicion that entanglement was not really real, nothing more than some exotic particles that live
for merely a tiny fraction of a second. But today we see a growing number of remarkable experiments
mastering entanglement. Entanglement over long distances
\cite{Tittel98,Tittel99,Weihs98,Marcikic04,Pan05}, entanglement between many photons \cite{Pan04}
and many ions \cite{Blatt05}, entanglement of an ion and a photon \cite{Monroe04,Weinfurter05},
entanglement of mesoscopic systems (more precisely entanglement between a few collective modes
carried by many particles) \cite{Polzik04,plasmon,Kimble05}, entanglement swapping
\cite{pan98,jennewein01,deRiedmatten05}, the transfer of entanglement between different carriers
\cite{Tanzilli05}, etc.

In summary: entanglement exists and is going to affect future technology. It is a radically new
concept, requiring new words and a new conceptual category.

\section{From quantum nonlocality to mere nonlocality}\label{nonLocality}
So far we have seen that quantum physics produces nonlocal correlations. And so what? Ok, this can
be used for Quantum Key Distribution and other Quantum Information processes, but that doesn't help
much to understand non-locality. Conceptually, one would like to study non-locality without all the
quantum physics infrastructure: Hilbert spaces, observables and tensor products. Not too
surprisingly, once the existence of non-locality was accepted, the conceptual tools to study it
came very naturally. Actually, the tools were already there, in the mathematics
\cite{TsirelsonMachine93} and even the physics \cite{PR94,PR9597} literature, waiting for a
community to wake up! The basic tool is simple, doesn't require any knowledge of quantum physics
and allows one, so to say, to study quantum nonlocality "from the outside", i.e. from outside the
quantum physics infrastructure.

Let us go back to the quantum exam \#3 (subsection \ref{QE3}). Assume that Alice and Bob are not
restricted by quantum physics, but only restricted by no-signaling. Consequently, they would fail
the quantum exam \#1. But under this mild no-signaling condition they could perfectly succeed in
the quantum exam \#3: Alice and Bob would each output a bit which locally looks perfectly random
and independent from their inputs - hence there would be no signaling - yet their 2 bits would
satisfy $a+b=x\cdot y$, exactly as in the coin tossing game of section \ref{CoinTossing}.
The hypothetical "machine" that produces precisely this correlation is a
basic example of the kind of conceptual tools we need to study nonlocality without quantum physics.
Formally, the correlation function is defined by:
\beq
P(a,b|x,y)=\half\delta(a+b=x\cdot y) \label{abxy}
\eeq
where the $\delta(z_1=z_2)$ function takes value 1 for $z_1=z_2$ and value 0 otherwise.

The correlation (\ref{abxy}) is often referred to as a PR-box, to recall the seminal work by
Popescu and Rohrlich \cite{PR94,PR9597}, or as a NL-machine (a Non-Local 
machine\footnotemark\footnotetext{A "machine" is a physicists's terminology for an input-output black-box
that is not necessarily mechanical. I believe that this terminology appeared in the quantum physics
context with the "optimal cloning machines" introduced by Buzek and Hillery
\cite{BuzekHillery96}}). The idea of these terminologies is to emphasize the similarities between
quantum measurements on 2 maximally entangled qubits and the correlation (\ref{abxy}): in both
cases the outcome is available as soon as the corresponding input is given (Alice knows $a$ as soon
as she inputs $x$ into her part of the machine and similarly Bob knows $b$ as soon as he inputs
$y$, there is no need to wait for the other's input) and in both the quantum and the PR-box cases
the "machine" can't be used more than once (once Alice has input $x$, she can't change her mind and
give another input). Notice a third nice analogy, neither the quantum nor the NL machines allow for
signaling. Indeed, in all cases the marginals are pure noise, independently of any input.

Note that quantum physics is unable to produce the PR correlation (\ref{abxy}). Indeed, this
correlation violates the Bell inequality (\ref{BellM}) up to its algebraic maximum, $M=4$, while
Tsirelson's theorem \cite{TsirelsonBound} states that quantum correlations are restricted to $M\le
2+\sqrt{2}$. However, the correlation (\ref{abxy}) is much simpler than quantum correlations, while
sharing many of their essential features. In particular (\ref{abxy}) is nonlocal but non-signaling.

In order to get some deeper understanding of the power of this hypothetical machine (\ref{abxy}) as
a conceptual tool, let us consider 3 properties of quantum correlations (many further nice aspects
can be found in \cite{MAG05,BLMPPR05,infres}). First we shall consider the so-called quantum
no-cloning theorem and see that it is actually not a quantum theorem, but a no-signaling theorem.
The next natural step is to analyze quantum cryptography, whose security is often said to be based
on the no-cloning theorem, and as we would expect by now, we shall find "non-signaling cryptography".
Finally, we consider the question of the communication cost to simulate maximal quantum
correlation. But before all this we need to recall some facts about non-signaling correlations.

\subsection{The set of non-signaling correlations}\label{noSignalPolytope}
Let us consider the set of all possible bi-partite correlations $P(a,b|x,y)$, where the inputs are
taken from finite alphabets $\{x\}$ and $\{y\}$ and similarly for the outputs  $\{a\}$ and $\{b\}$,
and which are non-signaling:
\beq
\sum_b P(a,b|x,y)=P(a|x) \mathrm{~is ~independent ~of~} y
\eeq
\beq
\sum_a P(a,b|x,y)=P(b|y) \mathrm{~is ~independent ~of~} x
\eeq
A priori this set looks huge. But it has a nice structure. First, it is a convex set: convex
combinations of non-signaling correlations are still non-signaling. Second, there are only a finite
number of extremal points (mathematician call such sets {\it polytopes} and the extremal point {\it
vertices}); accordingly every non-signaling correlation can be decomposed into a (not necessarily
unique) convex combination of extremal points. This is analog to quantum mixed states that can be
decomposed into convex mixtures of pure states.

Among the non-signaling correlations are the local ones, i.e. those of the form (\ref{lhv}), analog
to separable quantum states. The set of local correlations also forms a polytope, a
sub-polytope of the non-signaling one. Moreover all vertices of the local polytope are also
vertices of the non-signaling polytope, see Fig. 2 \cite{BLMPPR05}. The facets of the local
polytope are in one-to-one correspondence with all tight Bell inequality.

\begin{figure}
  \includegraphics[width=\columnwidth]{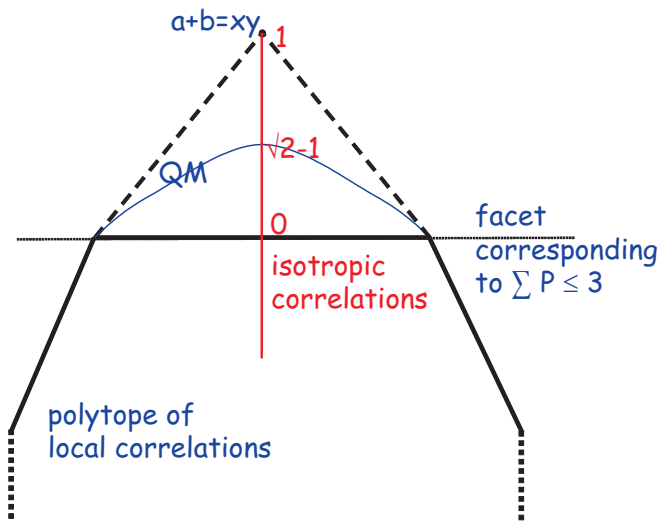}\\
  \caption{Geometrical view of the set of correlations. The bottom part represents the convex set (polytope)
  of local correlations, with the upper facet corresponding to the Bell inequality (\ref{BellM}). The upper
  triangle corresponds to the non-local non-signaling correlations that violate the Bell inequality. The smooth
  thin curve limits the correlations achievable by quantum physics. The top of the triangle corresponds to the
  unique non-signaling vertex above this Bell inequality, i.e. to the non-local PR machine (\ref{abxy}).
  The thin vertical line} represents the isotropic correlations (\ref{pNL}) with the indication of some of the values of
  $p_{NL}$.\label{FigPolytope}
\end{figure}

Let us illustrate this for the simple binary case (which is in any case the only one we need in this
article), i.e. $a,b,x,y$ are 4 bits. In this case, it is known that there are only 8 non-trivial
Bell inequalities (i.e. not counting the trivial inequalities of the form $P(a,b|x,y)\ge 0$), i.e.
only 8 relevant facets of the local polytope. Interestingly, Barret and co-workers \cite{BLMPPR05}
demonstrated that the non-signaling polytope has only 8 vertices more than the local polytope,
exactly one per Bell inequality! Each of these 8 vertices is equivalent to the PR correlation
(\ref{abxy}), up to an elementary symmetry (flip an input and/or an output). Although these
polytopes live in an 8-dimensional space\footnotemark\footnotetext{More precisely, 8 is the
dimension of the space of non-signaling correlations \cite{CollinsG04}.}, their essential
properties can be recalled from the simple geometry of figure 2.

\subsection{No-cloning theorem}\label{nocloning}
Details can be found in \cite{MAG05}, as here we would merely like to present the intuition. Let us assume that
Alice (input and output bits $x$ and $a$, respectively) shares the correlation (\ref{abxy}) both
with Bob (bits $y$ and $b$) and with Charly (bits $z$ and $c$): $a+b=xy$ and $a+c=xz$. Note that
this situation is different from the case where Alice would share one "machine" with Bob and share
another independent "machine" with Charly: in the situation under investigation Alice holds a
single input bit $x$ and a single output bit $a$. We shall see that the assumption that Alice's input
and output bits $x$ and $a$ are correlated both to Bob and to Charly leads to signaling. Hence in a
Universe without signaling, Alice can't share the correlation (\ref{abxy}) with more than one
partner: the correlation can't be cloned.

In order to understand this, assume that Bob and Charly come together, input $y=1$ and $z=0$, and
add their output bits $b+c$. According to the assumed correlations and using the modulo 2
arithmetic $a+a=0$, one gets: $b+c=a+b+a+c=xy+xz=x$. Hence, they could determine from their data
that Alice's input bit is $x$, i.e. Alice could signal to them!

A natural question is how noisy should the correlation (\ref{abxy}) be to allow cloning? The answer
is interesting: as long as the Alice-Bob correlation violates the Bell inequality (\ref{BellM}),
the Alice-Charly correlation can't violate it; if not there is signaling.

We have just seen that the CHSH-Bell inequality (\ref{BellM}) is monogamous, like well kept
secrets. Let's now see that this is not a coincidence!

\subsection{Non-signaling cryptography}\label{nonlocality}
In 1991 Artur Ekert's discovery of quantum cryptography \cite{Ekert91} based on the violation of
Bell's inequality changed the (physicist's) world: entanglement and quantum nonlocality became
respectable. Now, as we shall see in this subsection, the essence of the security of quantum
cryptography does not come from the Hilbert space structure of quantum physics (i.e. not from
entanglement), but is due to no-signaling nonlocal correlation! The fact that quantum physics
offers a way to realize such correlation makes the idea practical. However, if one would find any
other way to establish such no-signaling nonlocal correlations (a way totally unknown today), then
this would equally well serve as a mean to establish cryptographic keys \cite{AGM05}.

Let us emphasize that the goal is to assume no restriction on the adversary's power, i.e. on Eve,
except no-signaling\footnote{No-signaling should be understood here as in the previous sub-section on the no-cloning theorem.
That is, even if two parties joint, for example Eve and Bob come together, then they should not be able to
infer any information about the third party's input, e.g. Eve and Bob should not have access no Alice's input.} \cite{BHK05}. Obviously, if one assumes additional restrictions on Eve, like
restricting her to quantum physics, then Alice and Bob can distill more secret bits from their
data. But qualitatively, the situation would remain unchanged.

Assume that two partners, Alice and Bob, hold devices that allow them to each input a bit (make a
binary choice of what to do, e.g. which experiment to perform) and each receives an output bit
(e.g. a measurement result). This can be cast into the form of an arbitrary correlation:
$P(a,b|x,y)$, with $a,b,x,y$ four bits. Assume furthermore that the devices held by Alice and Bob
do not allow signaling. This simple and very natural assumption suffices to give a nice structure to
the set of correlations $P(a,b|x,y)$: as we recall in subsection \ref{noSignalPolytope} this set is
convex and has a finite number of extreme points, called vertices. The nice property is that any
correlation $P(a,b|x,y)$ can be decomposed into a convex combination of vertices, hence once one
knows the vertices one knows essentially everything. If the correlation is local, i.e. of the form
(\ref{lhv}), then it is not useful for cryptography; indeed the adversary Eve may know the strategy
$\lambda$. Hence, let's assume that $P(a,b|x,y)$ violates the Bell inequality (\ref{BellM}).
Consequently $P(a,b|x,y)$ lies in a well defined corner of the general polytope, a sub-polytope.
Barrett and co-workers found that this sub-polytope has only 9 vertices \cite{BLMPPR05}, 8 local
ones for which $M=3$ and only one nonlocal vertex, that corresponding to our conceptual tool, i.e.
to $a+b=xy$, for which $M=4$, see Fig. 2.

 In the case that Alice and
Bob are maximally correlated (maximally but non-signaling!), i.e. their correlation correspond to
the nonlocal vertex of Fig. 2, it is intuitively clear that the adversary Eve can't be correlated
neither to Alice, nor to Bob, by the no-cloning argument sketched in the previous subsection.
Hence, in such a case Alice and Bob receive from their apparatuses perfectly secret bits. However,
these bits are not always correlated: when $x=y=1$ they are anti-correlated. But this can be easily
fixed by the following protocol. After Alice and Bob received their output bits, Alice announces
publicly her input bit $x$ and Bob changes his output bit to $b'=b+xy$. Now Alice and Bob are
perfectly correlated and Eve still knows nothing about $a$ and $b'$.

Consider now that Alice and Bob are non maximally correlated:
\beq
P(a,b|x,y)=\frac{1+p_{NL}}{2}\half\delta(a+b=x\cdot y) + \frac{1-p_{NL}}{2}\frac{1}{4} \label{pNL}
\eeq
For $p_{NL}>0$ this correlation violates the inequality (\ref{BellM}), for $p_{NL}\le \sqrt{2}-1$,
it can be realized by quantum physics. Can Alice and Bob exploit such a correlation for
cryptographic usage secure against an arbitrary adversary who is not restricted by quantum physics,
but only restricted by the no-signaling physics? The full answer to this fascinating question is
still unknown. However, there is an optimistic answer if one assumes that Eve attacks each
realization independently of the others, the so-called individual attacks. In such a case, one may
assume that Eve does actually distribute the apparatuses to Alice and Bob. Some apparatuses are
ordinary local ones, for these Eve knows exactly the relation between the input and output bits,
both for Alice and for Bob. For example, Eve sends to Alice an apparatus that always outputs a 0,
and to Bob an apparatus that outputs the input bit: $b=y$. In this example Eve knows Alice's bit
$a$, but doesn't know Bob's bit. For some local pairs of apparatuses Eve knows both $a$ and $b$, or
$b$ but not $a$. But, if the Alice-Bob correlation (\ref{pNL}) violates the Bell inequality
(\ref{BellM}), i.e. if $p_{NL}>0$, then Eve must sometimes send to Alice and Bob the apparatuses
that produce the maximal nonlocal correlation $a+b=xy$ \footnotemark\footnotetext{One may think
that Eve should sometimes send a weakly non-local machine. But all such correlations are convex
combinations of local and fully non-local NL-machines. Hence, it is equivalent for Eve to always
send either a local or a NL-machine, with appropriate probabilities.}, in which case she knows
nothing about Alice and Bob's output bits $a$ and $b$. A detailed analysis can be found in
\cite{AGM05}. Here we merely recall the result. For $p_{NL}>0.318$ the Shannon mutual information
between Alice and Bob is larger than the Eve-Bob mutual information \cite{AGM05}. Hence for
$p_{NL}>0.318$ Alice and Bob can distil a cryptographic secret key out of their data, secure even
against an hypothetical post-quantum adversary, provided this adversary is still subject to
no-signaling.

Actually, in \cite{AGM05} we worked out a 2-way protocol for key distillation valid down to $p_{NL}>0.09$.
There, it is also proven that the intrinsic information is positive for all positive $p_{NL}$. It is thus tempting to conjecture that secret key distillation is possible if and only if the Bell inequality is violated\footnotemark\footnotetext{In \cite{MAG05} we proved that a correlation $P(a,b|x,y)$ is nonlocal
iff any possible non-signaling extensions $P(a,b,e|x,y,z)$ has positive Alice-Bob condition mutual
information, conditioned on Eve, $I(A,B|E)$, i.e. has positive intrinsic information. This nicely
complements the similar result that holds for entangled quantum states and purifications
\cite{intrInfo}. In \cite{AGM05} we proved that the same relation between nonlocality and positive
intrinsic information does also hold when Alice announces her input and Bob adapts his output in
such a way as to maximize his mutual information with Alice. Proving this in full generality would
be a marvellous result.}.


Another beautiful result is the observation of an {\it information gain versus disturbance}
relationship, very similar to that of quantum physics, based on Heisenberg's uncertainty relations
\cite{Scarani05}. Let us analyze separately the cases where Alice announces $x=0$ and $x=1$, and
denote the respective Alice-Bob error rates $QBER_x$ and the Eve-Bob mutual informations
$I_x(B,E)$, i.e. $QBER_x=\sum_yP(a\ne b)|x,y)$ and $I_x(B,E)=H(B|x)-H(B|E,x)$. Remarkably,
$I_0(B,E)$ is a function of only $QBER_1$ and $I_1(B,E)$ of $QBER_0$
\footnotemark\footnotetext{Precisely one has: $I_0(B,E)=2\cdot QBER_1$ and $I_1(B,E)=2\cdot
QBER_0$.}: information gain for one input necessarily produces errors for the other input, in
analogy with the quantum case where information gain on on basis necessarily perturbs information
encoded in a conjugated basis!

To conclude this subsection, let us emphasize that the distribution of the correlation (\ref{pNL})
by quantum means requires a protocol that differs from the famous BB84 protocol \cite{BB84}.
Indeed, the data obtained by Alice and Bob following the BB84 protocol do not violate any Bell
inequality, hence the BB84 protocol is not secure against a non-signaling post-quantum adversary.
Indeed, even the noise-free BB84 data can be obtained from quantum measurements on a separable
state in higher dimension. The additional dimension could, for the example of polarization coding,
be side-channels due to accidental additional wavelength coding. Consequently, standard security
proofs \cite{ShPr00,KrausRG05} must make assumptions about the dimension of the relevant Hilbert
spaces (accordingly, no security proof of quantum key distribution is unconditional, contrary to
widespread claims). But it is easy to adapt the BB84 protocol, it suffices that Alice measures the physical quantities corresponding to the Pauli matrices
$\sigma_z$ or $\sigma_x$, depending on her input bit value 0 or 1, respectively, exactly as in
BB84, and Bob measures in the diagonal bases: $\sigma_{+45^o}$ and $\sigma_{-45^o}$ for $y=0$ and
$y=1$, respectively. In this way Alice and Bob's data are never perfectly correlated, but they can
violate the Bell inequality and be thus exploited to distil a secret key valid even against
post-quantum adversaries. Note that the violation of a Bell inequality guarantees that no side channels
accidentally leak out information. Furthermore, in this protocol, that we like to call the CHSH-protocol, in
honor of the 4 inventors \cite{CHSH} of the most useful Bell inequality (actually equivalent to
(\ref{BellM})), Alice announces her input bit $x$, i.e. her basis as in BB84, but Bob doesn't
speak, he always accepts and merely flips his bit in case $x=y=1$. In summary, in the CHSH protocol
Alice and Bob use all the raw bits, however their data are initially noisier than in the BB84
protocol.

\subsection{Cost of simulating quantum correlations}\label{simulation}
Among the many contributions of computer science to quantum information is the beautifully simple
question (actually anticipated by Maudlin \cite{Maudlin92}): what is the cost of simulating quantum
correlations? More precisely, Gilles Brassard, Richard Cleve and their student Alain Tapp
\cite{TappCB99}, and independently Michael Steiner \cite{Steiner00}, asked the question: {\it How
many bits must Alice and Bob exchange in order to simulate (projective) measurement outcomes
performed on quantum systems?} The question concerns the communication during the measurement
simulation, clearly there must have been prior agreement on a common strategy. If the systems are
in a separable state, no communication at all is needed. On the contrary, if the state allows
measurements that violate a Bell inequality, i.e. if the state has quantum nonlocality, then it is
impossible to simulate it without some communication or some other nonlocal resources.

For the simplest case of two 2-level systems (2 qubits), this game assumes that Alice and Bob
receive as input any possible observable, i.e. any vector $\vec a$ and $\vec b$ of the Poincar\'e
sphere. And they should output one bit, corresponding to the binary measurement outcome "up" or
"down" in the physicist's spin $\half$ language. A simple way to simulate the quantum measurements
is that Alice communicates her input $\vec a$ to Bob and outputs a predetermined bit (predetermined
by Alice and Bob's common strategy).
But communicating a vector corresponds to infinitely many bits! My first intuition was that there
is no way to do any better, after all the input space is a continuum, quite the contrary to the
case of Bell inequalities where the input space is finite, usually even limited to a binary choice.
Yet, Brassard and co-workers came out with a model using only 8 bits of communication! What a
surprise: is entanglement that cheap? But this was only a start. Steiner published a model valid
only for vectors lying on the equator of the sphere, but this model was easy to generalize to the
entire sphere \cite{GisinG99}: it uses only 2 bits! 2 bits, like in dense coding and teleportation:
that should be the end, I thought! But, yet again, I was wrong. Bacon and Toner produced a model
using one single bit of communication \cite{TonerB03}. Well, by now we should be at the limit,
isn't it? But actually, not quite!

Let's come back to the real central question: How does Nature manage to produce random data at
space-like separated locations that can't be explained by common causes? The idea that Nature might
be exploiting some hidden communication (hidden to us, humans) is interesting. With my group at
Geneva University we spent quite some time trying to explore this idea, both experimentally and
theoretically. We could set experimental bounds of the speed of this hypothetical hidden
communication \cite{SpeedQI}. We also investigated the idea that each observer sends out hidden
information about his result at arbitrary large speeds as defined in its own inertial reference
frame \cite{MovingObs}. The measured bounds on the speed of the hypothetical hidden communication
were very high and the latter assumption contradicted
by experiments. Also our theoretical investigation cast serious doubts on the existence of hidden
communication. Analyzing scenarios involving 3 parties we could prove that if all quantum
correlations would be due to hidden communication, then one should be able to signal (i.e. the
hidden communication do not remain hidden) \cite{ScaraniG02,ScaraniG05}! Hence, the only remaining
alternative is that Nature exploits both hidden communication and hidden variables: each one
separately contradicts quantum theory, but both together could explain quantum physics. However,
this seems quite an artificial construction. Hence, let's face the situation: Nature is able to
produce nonlocal data without any sort of communication. But is she doing so using all the quantum
physics artillery? Aren't there logical building blocks of nonlocality? A partial answer follows.

Let us come back to the problem of simulating quantum measurements, but instead of a few bits of
communication let us give Alice and Bob a weaker resource: one instance of the nonlocal machine
$a+b=xy$. That this is indeed a weaker resource follows from the observation that the correlation
$a+b=xy$ can't be used to communicate any bit, but that by sending a single bit one can easily
simulate the nonlocal correlation (since Alice's input is only a bit $x$, it suffices that she
communicates it to Bob). The nice surprise is that this elementary resource is sufficient to simulate any
pair of projective measurements on any maximally entangled state of two qubits! For the proof the
reader is referred to the original article \cite{CGMP05} and to the beautiful account in
\cite{JulienJeremieS05} where the relations between all these models are presented.

The above results are very encouraging. One can gets the feeling that, at last, one can start understanding
nonlocality without the Hilbert space machinery, that, at last, one can study quantum physics from
the outside, i.e. from the perspective of future physical theories (assuming these will keep
Einstein's no-signaling constraint) and no longer from the perspective of the old classical mechanical physics.
But there is still a lot to be done! For instance, it is surprising (and annoying in my opinion)
that one is still unable to simulate measurement on partially entangled states using the nonlocal
correlation (actually we could prove that this is impossible with a single instance of the
correlation, but there is hope that one can simulate partially entangled qubit pairs with 2
instances \cite{BrunnerSG05}). Let me emphasize that all of today's known simulation models for
partially entangled qubits include some sort of communication\footnotemark\footnotetext{Using the reduction of an OT-box (Oblivious Transfer to a PR-box) \cite{WolfW05}
one can simulate any 2-qubit state with one OT-box.} \cite{TonerB03}, let's say
from Alice to Bob. Consequently, in all these models Bob can't output his results before Alice was
given her input. This contrasts with the situation in quantum measurements where Bob doesn't need to
wait for Alice (he does not even need to know about the existence of Alice) and with the simulation
model for maximally entangled qubits using the PR-box. It would be astonishing if partially
entangled state could not be simulated in a time-symmetric way \cite{PartialEnt}.

\section{Conclusion}\label{concl}
The history of non-locality in physics is fascinating. It goes back to Newton (section \ref{Newton}.
It first accelerated around 1935 with
Einstein's EPR and Schr\"odinger cat's papers. Next, it slowly evolved, with the works of John
Bell, John Clauser and Alain Aspect among many others, from a mere philosophical debate to an
experimental physics question, or even to {\it experimental metaphysics} as Abner Shimony nicely
put it \cite{ExpMeta}. Now, during the last decade, it has run at full speed. Conceptually the two major
breakthroughs were, first Artur Ekert's 1991 PRL which strongly suggests a deep link between
non-locality and cryptography, section \ref{nonlocality}. The second breakthrough, in my opinion,
is the PR-box, section \ref{noSignalPolytope}, the
understanding that non-signaling correlations can be analyzed for themselves, without the need of
the usual Hilbert space artillery, thus providing a simple conceptual tool for the unravelling of
quantum non-locality. We have reviewed that the no-cloning theorem, the uncertainty relation, the
monogamy of extreme correlation and the security of key distribution, all properties usually
associated to quantum physics are actually properties of any theory without signaling, section \ref{nonLocality}. In
particular we emphasized that the second breakthrough, the PR-box, allows one to confirm the first
breakthrough: there is an intimate connection between violation of a Bell inequality and security of
quantum cryptography.

And relativity, can it be considered complete? Well, if nonlocality is really real, as widely
supported by the accounts summaries in this article, then all complete theories should have a place
for it. Hence, the question is: "Does relativity hold a place for non-signaling nonlocal
correlations?".

\small
\section*{Acknowledgment}
This article has been inspired by talks I gave in 2005 at the IOP conference on Einstein in Warwick, 
the QUPON conference in Vienna, the {\it Annus Mirabilis} Symposium in Zurich, le s\'eminaire de l'Observatoire de Paris
and the Ehrenfest Colloquium in Leiden. 
This work has been supported by the EC under projects RESQ and QAP (contract n.
IST-2001-37559 and IST-015848) and by the Swiss NCCR {\it Quantum Photonics}.


\end{document}